\documentclass[a4paper,fleqn,usenatbib]{mnras}
\usepackage{graphicx}
\usepackage[T1]{fontenc}
\usepackage{ae,aecompl}

\usepackage{amsmath}	% Advanced maths commands
\usepackage{amssymb}	% Extra maths symbols

\usepackage{url}

\hypersetup{draft}

\usepackage{upgreek}
\usepackage[usenames]{color} 
\usepackage{appendix}
\usepackage{color}
\usepackage{float}

\title[HD~12098]{HD~12098: a highly distorted dipole mode in an obliquely pulsating roAp star}

\author[D.~W.~Kurtz et al.]{D.~W.~Kurtz,$^{1,2}$\thanks{E-mail: kurtzdw@gmail.com} H.~Saio,$^3$ D.~L.~Holdsworth,$^{2,4}$ Santosh~Joshi,$^5$ and S.~Seetha$^6$\\
$^{1}$Centre for Space Research, North-West University, Dr Albert Luthuli Drive, Mahikeng 2735, South Africa\\
$^{2}$Jeremiah Horrocks Institute, University of Central Lancashire, Preston PR1 2HE, UK\\
$^{3}$Astronomical Institute, Graduate School of Science, Tohoku University, Sendai 980-8578, Japan\\
$^{4}$South African Astronomical Observatory, P.O. Box 9, Observatory 7935, Cape Town, South Africa\\
$^{5}$Aryabhatta Research Institute of Observational Sciences, Nainital 263001, Uttarakhand, India\\
$^{6}$Raman Research Institute, C.V. Raman Avenue, Bengaluru, 560080, India\\
}

\date{\today}
\pagerange{\pageref{firstpage}--\pageref{lastpage}} \pubyear{2024}
\begin{document}
\maketitle
\label{firstpage}

\begin{abstract}
HD~12098 is an roAp star pulsating in the most distorted dipole mode yet observed in this class of star. Using {\it TESS} Sector 58 observations we show that there are photometric spots at both the magnetic poles of this star. It pulsates obliquely primarily in a strongly distorted dipole mode with a period of $P_{\rm puls} = 7.85$\,min ($\nu_{\rm puls} = 183.34905$\,d$^{-1}$; 2.12210\,mHz) that gives rise to an unusual quadruplet in the amplitude spectrum. Our magnetic pulsation model cannot account for the strong distortion of the pulsation in one hemisphere, although it is successful in the other hemisphere. There are high-overtone p~modes with frequencies separated by more than the large separation, a challenging problem in mode selection. The mode frequencies observed in the {\it TESS} data are in the same frequency range as those previously observed in ground-based Johnson $B$ data, but are not for the same modes. Hence the star has either changed modes, or observations at different atmospheric depth detect different modes. There is also a low-overtone p~mode and possibly g~modes that are not expected theoretically with the $> 1$\,kG magnetic field observed in this star.
\end{abstract}  


\section{Introduction}

Magnetic peculiar A (Ap) stars are found in the main-sequence band from spectral types early-B to mid-F. They have global magnetic fields that are roughly dipolar with strengths up 34\,kG and with a magnetic axis that is inclined to the rotation axis, so that the field is observed from varying aspect with rotation. Atomic diffusion gives rise to surface abundance anomalies that can reach a million times that of normal stars for some rare earth elements. Those enhanced abundances are concentrated in surface patches, usually referred to as spots, that are closely associated with the magnetic poles, giving rise to rotational variations in the spectral line strengths (hence abundances) and in brightness. These stars are known as $\alpha^2$\,CVn stars; they constitute about 10\,per~cent of all main-sequence stars in that spectral type range. 

A notable characteristic of the $\alpha^2$\,CVn stars is that the surface spots are stable over time-scales of at least decades, in strong contrast to lower main-sequence spotted stars for which those spots can evolve on time-scales of days. The Sun is the prime example of this. This stability in the Ap stars means that the rotational light curves can be used to determine the rotation period to high precision. While the spots are stable, they have a wide variety of surface configurations, often giving rise to non-sinusoidal light curves. When the rotational inclination, $i$,  and the magnetic obliquity, $\beta$, of the dipolar field sum to greater than 90$^{\circ}$, both magnetic poles are seen; also, when there are spots associated with both poles, the rotational light curve has a double-wave. 

The rapidly oscillating Ap (roAp) stars are a subset of the magnetic Ap stars that show high radial overtone p~mode pulsation with periods in the range $4.7 - 23.6$\,min (see Holdsworth et al., 2024, in press). The spectral types of the roAp stars range from early A to late F with effective temperatures of $6500 \le T_{\rm eff} \le 8800$\,K. They partially overlap with the $\delta$~Sct stars in the HR~Diagram, but differ from those because of the impact of the magnetic field on the pulsations. The roAp stars generally pulsate in nonradial, zonal dipole and quadrupole modes ($l = 1,2; m=0$) with the pulsation axis lying close to the magnetic axis. This gives rise to oblique pulsation where the pulsation mode is seen from varying aspect as the star rotates. That provides information on the mode geometry, hence constrains mode identification, which in many other stars can be problematic. Mode identification is requisite for the application of asteroseismic modelling. 

The oblique pulsator model was introduced by \citet{1982MNRAS.200..807K} and further developed and improved by \citet{1985PASJ...37..245S}, \citet{1993PASJ...45..617S}, \citet{1994PASJ...46..301T, 1995PASJ...47..219T}, \citet{2002A&A...391..235B} and \citet{2011A&A...536A..73B}, among others. For pulsation modes that can be described by single spherical harmonics, the pulsation amplitude and phase changes that are observed over the rotation cycle allow the oblique pulsator model to give constraints on $i$ and $\beta$, thus on the pulsation geometry, hence identifying the mode. Two well-studied cases, HR~3831 and HD~99563, pulsate in slightly distorted dipole modes for which the rotational pulsation amplitude and phase modulation provide good constraints on the pulsation geometry (\citealt{1990MNRAS.247..558K}, \citealt{2006MNRAS.366..257H}).

Observations show that in other roAp stars the modes are often more strongly distorted. \citet{2005MNRAS.360.1022S} made a nonadiabatic analysis of nonradial pulsation modes in the presence of a dipole field for roAp stars, finding that the dipole and quadrupole modes distorted by the magnetic field are most likely to be excited. Several cases of strongly distorted quadrupole modes were studied by \citet[][and references therein]{2018MNRAS.476..601H} and modelled with the method of \citet{2005MNRAS.360.1022S}. This was particularly successful in explaining the flattening of the pulsation phase curve as a function of rotation phase. 

However, the models are not fully capable of explaining the distortion of the pulsation modes, and further complications and challenges to the oblique pulsator model, as applied to the roAp stars, have arisen. In particular, \citet{2020ASSP...57..313K} showed that the geometry deduced for the dipole pulsation mode in the roAp star HD~6532 is dramatically different for observations made in Johnson $B$ and for those made with the red-dominated bandpass of the Transiting Exoplanet Survey Satellite ({\it TESS}) mission. Since observations at these different wavelengths probe to different atmospheric depths because of differences in opacity, this suggests strong changes in pulsation geometry as a function of depth. In a high-resolution spectroscopic study of the distorted dipole mode in HR~3831, \citet{2006A&A...446.1051K} first pointed out that the pulsation geometry inferred using the oblique pulsator model is dependent on the depth of the observations. 

Hence the roAp stars continue to present challenges to our understanding of their oblique pulsation. This is particularly true for the stars with the most distorted mode geometries. This has recently become relevant in the context of the tidally tilted pulsators, where tidally distorted zonal and sectoral dipole modes have been observed (\citealt{2022ApJ...928L..14J, 2021MNRAS.503..254R, 2020MNRAS.498.5730F, 2020MNRAS.494.5118K, 2020NatAs...4..684H}). Thus the study of oblique pulsation of distorted nonradial modes has expanding applications. Therefore, in this work we present the most distorted dipole mode yet observed in an roAp star. Extreme examples provide the strongest challenges to theory. 

\subsection{HD~12098}

HD~12098 has only one published spectral classification, which is F0 from the original HD catalogue. However, its Str\"omgren colours and indices are characteristic of cool Ap stars; $b-y = 0.191$, $m_1 = 0.328$, $c_1 = 0.517$, H$\beta = 2.796$  \citep{1983A&AS...54...55O}, from which the indices $\delta m_1 = -0.122$ and $\delta c_1 = -0.255$ can be calculated from the calibration of \citet{1979AJ.....84.1858C}. Based on these indices, in 1999 the star was tested for pulsation under the Nainital-Cape Survey project (\citealt{2000BASI...28..251A}, \citealt{2001A&A...371.1048M}, \citealt{2006A&A...455..303J}, 2009, 2016) %\citealt{2009A&A...507.1763J}, \citealt{2016A&A...590A.116J}).
using 10-s integrations through a Johnson $B$ filter on the 1.04-m Sampurnanand  telescope of the Aryabhatta Research Institute of Observational Sciences (ARIES; at that time called the Uttar Pradesh State Observatory). \citet{2000IBVS.4853....1M} found it to be an roAp star with a pulsation frequency of 189.2\,d$^{-1}$ (2.19\,mHz; $P = 7.6$\,min).

\citet{2001A&A...380..142G} then obtained further data from Mt. Abu Observatory, India, and found the star to be multiperiodic with a dominant frequency of 187.82\,d$^{-1}$ (2.17\,mHz). They also found rotational variation, but could not discriminate between periods of 1.2\,d and 5.5\,d. That ambiguity was addressed by \citet{2001APN....35.....W} and then resolved by \citet{2005A&A...429L..55R} who found a rotation period of $5.460 \pm 0.001$\,d from a study of the mean longitudinal field, which varies almost sinusoidally between $+2000$\,G and $-500$\,G with that period. That shows that both magnetic poles are seen over the rotation cycle.

\section{HD~12098 data and analysis}

HD~12098 was observed at 120-s cadence by the {\it TESS} mission in Sector 58 in late 2022. We have used the PDCSAP (Presearch-Data Conditioning Simple Aperture Photometry) data to study the rotation and pulsation variations of this star. The data span 27.72\,d and comprise 19475 data points. No outliers were removed.  

\subsection{The rotation period}

\begin{figure}
\begin{center}
\includegraphics[width=1.0\linewidth,angle=0]{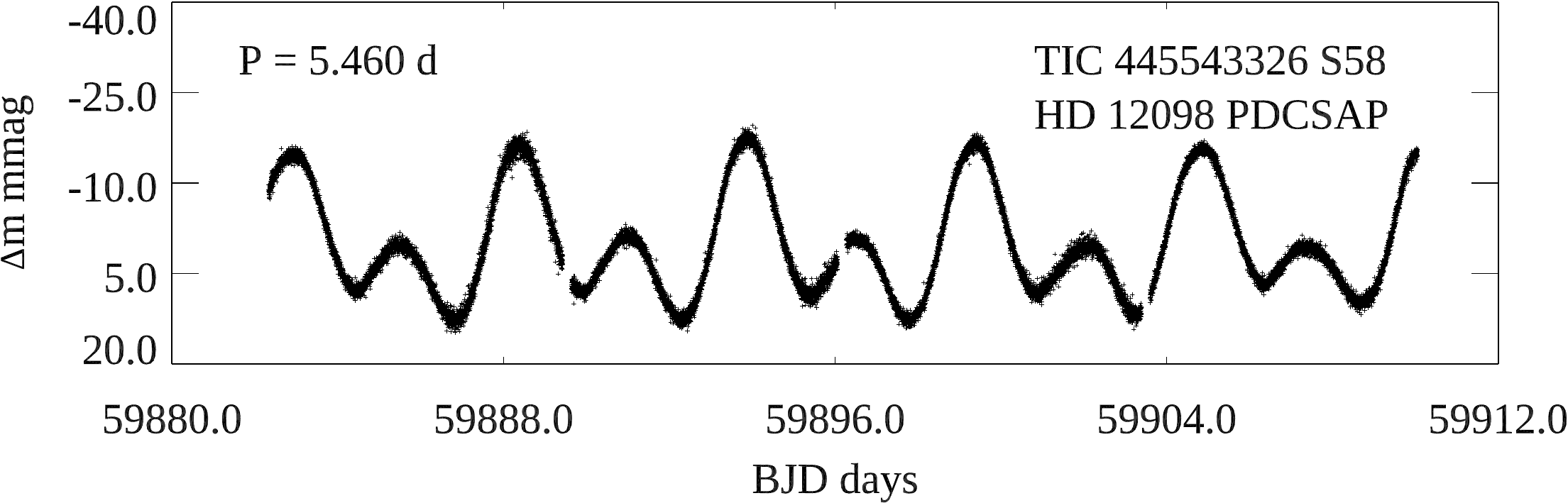}
\includegraphics[width=1.0\linewidth,angle=0]{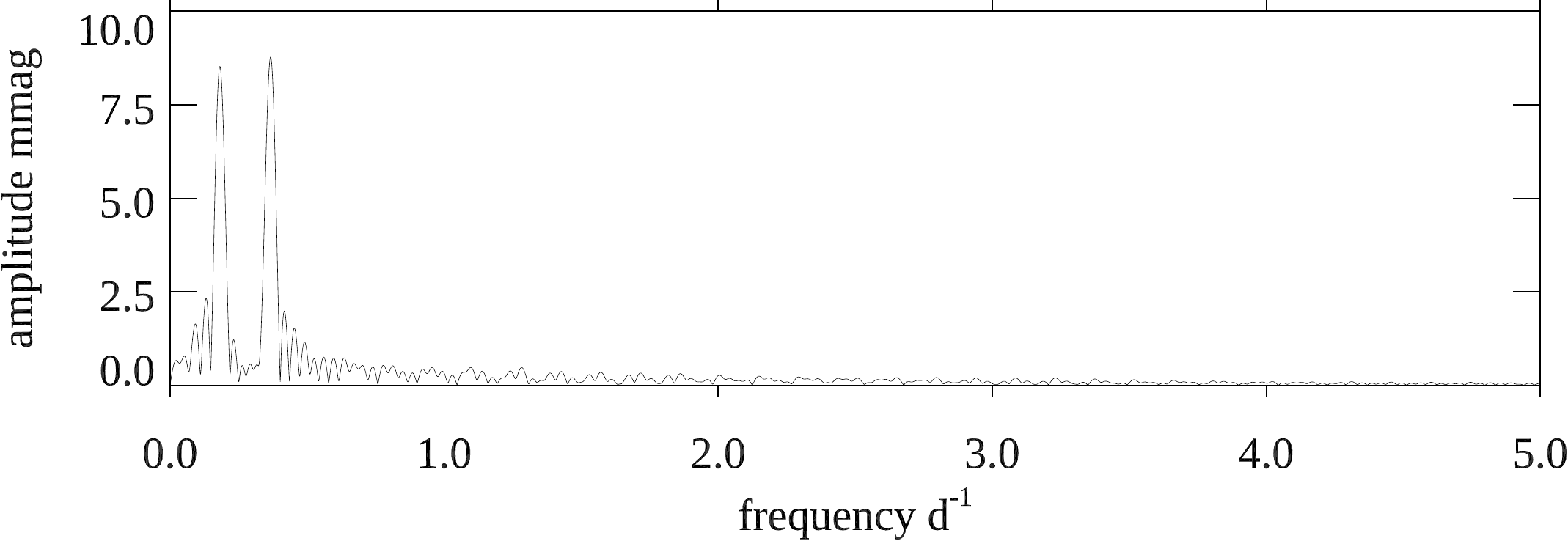}
\includegraphics[width=1.0\linewidth,angle=0]{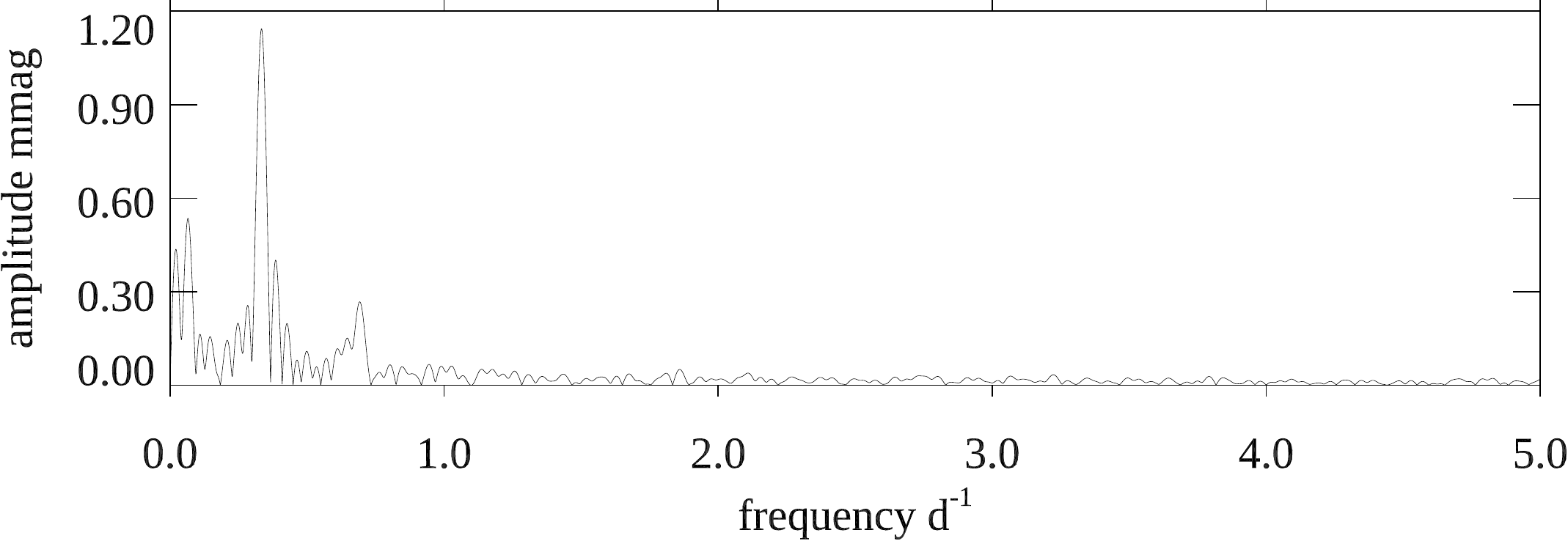}
\includegraphics[width=1.0\linewidth,angle=0]{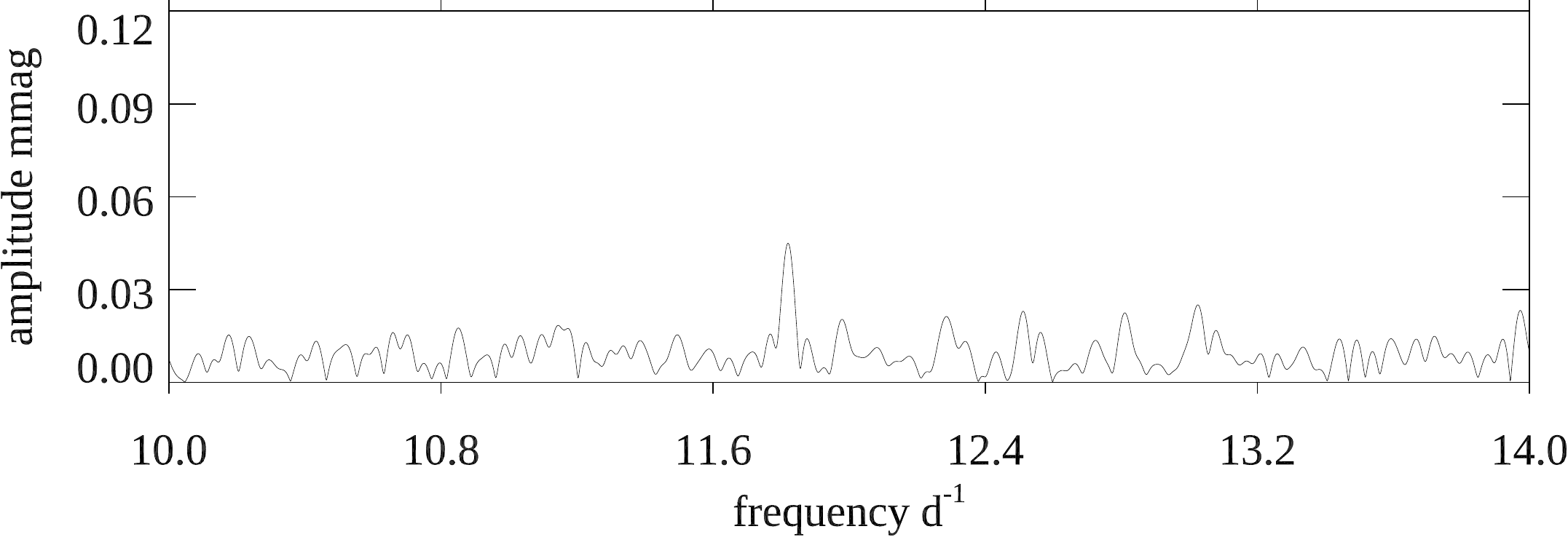}
\caption{Panel 1: The Sector 58 {\it TESS} light curve for HD~12098 showing the double-wave rotational variation typical of the $\alpha^2$\,CVn stars. Panel 2: A low-frequency amplitude spectrum showing the first two harmonics of the rotation frequency. Panel 3: A low-frequency amplitude spectrum after pre-whitening a 10-harmonic fit for the rotation frequency. There are significant peaks visible that may be from g~modes. One of these is not fully resolved from the second harmonic of the rotation frequency. Panel 4 shows a 5$\sigma$ peak that is potentially from a low-order p~mode. Note the changes of ordinate scale on the panels.}
\label{fig:lc-ft} 
\end{center}
\end{figure}

Fig.\,\ref{fig:lc-ft} shows the Sector 58 light curve in the top panel showing a clear double wave. This is consistent with polar spots and the longitudinal magnetic field curve of  \citet{2005A&A...429L..55R}. The second panel shows the first two harmonics of the rotational light variation in an amplitude spectrum calculated using the Discrete Fourier Transform algorithm of \citet{1985MNRAS.213..773K}. The rotation period determined from those harmonics is $ 5.4815 \pm 0.0002$\,d, which is significantly different to that determined from the magnetic variation by  \citet{2005A&A...429L..55R} of $ 5.460 \pm 0.001$\,d.

After fitting 10 harmonics of the rotation frequency, the third panel of Fig.\,\ref{fig:lc-ft} shows significant peaks that may indicate some g~modes. One of these is not fully resolved from the second harmonic of the rotation frequency, hence may have perturbed the value of the rotation period determined. Panel 4 shows a 5$\sigma$ peak at 11.819\,d$^{-1}$ that is potentially from a low-overtone p~mode. Because of this possible perturbation of the rotation frequency by an unresolved g~mode frequency, we adopt the rotation period (frequency) of $P_{\rm rot} = 5.460$\,d  ($\nu_{\rm rot} = 0.18315$\,d$^{-1}$) of \citet{2005A&A...429L..55R} for the pulsation analysis. Pre-whitening by a 10-harmonic series with this rotation frequency provides a good fit to the data.

\subsection{The pulsations} 

\begin{figure}
\begin{center}
\includegraphics[width=1.0\linewidth,angle=0]{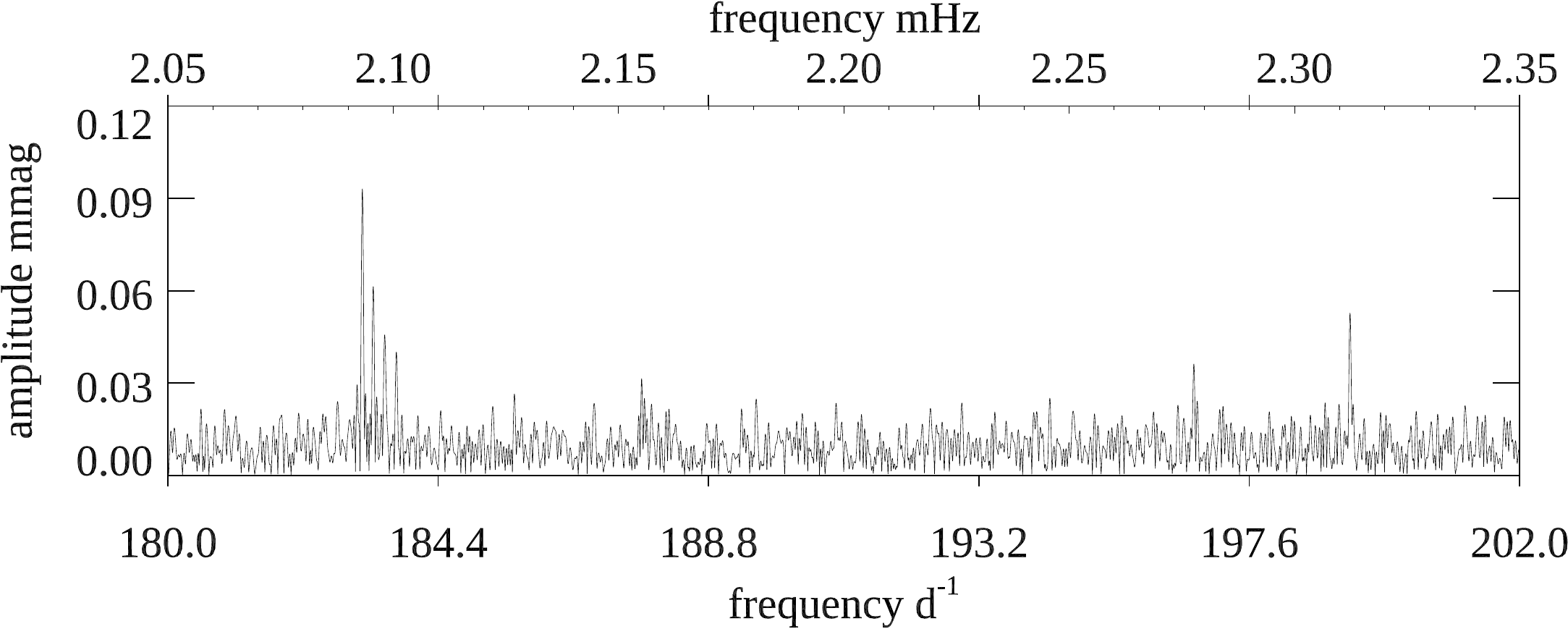}
\includegraphics[width=1.0\linewidth,angle=0]{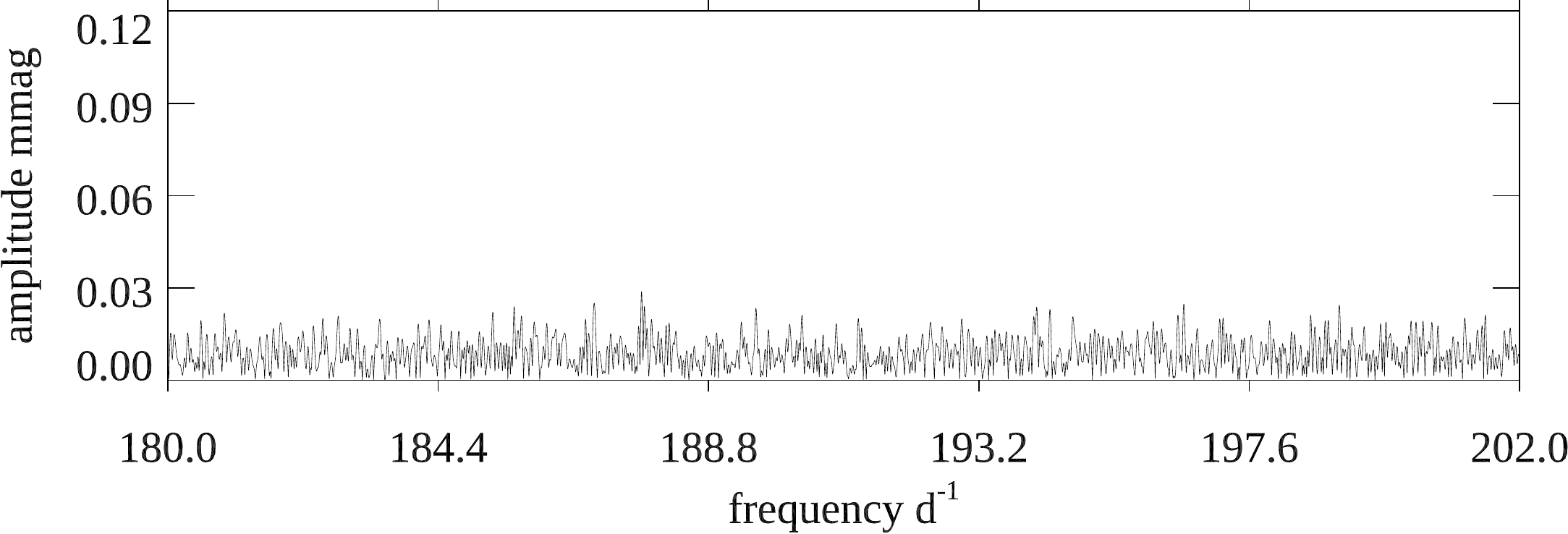}
\includegraphics[width=1.0\linewidth,angle=0]{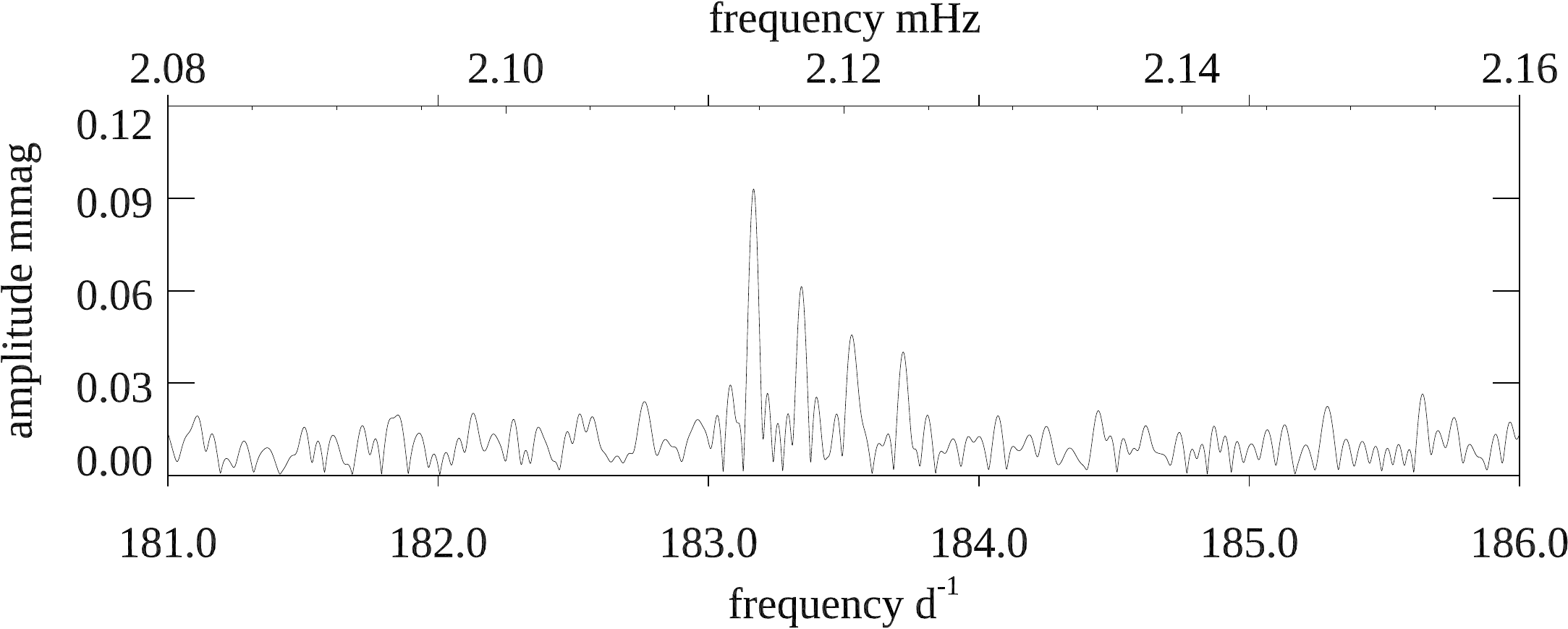}
\caption{Panel 1: The roAp pulsation frequencies. Panel 2: same as panel 1 after pre-whitening the 6 pulsation frequencies. Panel 3: a higher resolution view of the 183.3491\,d$^{-1}$ (2.1221\,mHz) rotational multiplet. }
\label{fig:lc-ft2} 
\end{center}
\end{figure}

A high-pass filter was used to remove the low-frequency rotational frequency and its harmonics, the (probable) g~mode and p~mode frequencies, and instrumental variations so that the noise in the amplitude spectrum is white. This gives the best uncertainty estimates for the derived roAp pulsation frequencies, amplitudes and phases. The top panel of Fig.\,\ref{fig:lc-ft2} shows a section of the amplitude spectrum in the range of the roAp pulsation frequencies where 6 significant peaks can be seen, four of which form a quadruplet split by exactly the rotation frequency. The middle panel shows the amplitude spectrum of the residuals after those 6 frequencies were pre-whitened. The bottom panel shows a higher resolution view of the quadruplet. Table\,\ref{Tab:ls} gives the frequencies derived from a least-squares fits of the quadruplet with a forced splitting equal to the rotation frequency, and the other two pulsation frequencies. 

Fig.\,\ref{fig:phase1} shows the rotational light variations and pulsation amplitude and phase as a function of rotation phase. The time zero point, $t_0 = {\rm BJD} \,245\,9893.80$, was chosen to be the time of maximum rotational brightness. The pulsation amplitude and phase were calculated from the frequency quadruplet, taking $\nu_2 = 183.34905$\,d$^{-1}$ (2.1221\,mHz) to be the pulsation frequency, and the other three frequencies, $\nu_1$, $\nu_3$, and $\nu_4$ to be rotational sidelobes generated by oblique pulsation. A noise-free time series was generated for this frequency quadruplet as given in Table\,1, sampled at the times of the actual observations. The pulsation amplitudes and phases were then derived by least-squares fitting of the pulsation mode frequency, $\nu_2$, to sections of the generated data $0.05$-d long, which is just over 9 pulsation cycles. Those were then plotted as a function of rotation phase. The choice of $\nu_1$ or $\nu_3$ as the pulsation frequency generated pulsation phase plots, such as the bottom panel of Fig.\,\ref{fig:phase1}, that had strong slopes, indicating that those are not the pulsation frequency. 

\subsection{Discussion of the pulsation frequencies}

\begin{figure}
\begin{center}
\includegraphics[width=1.0\linewidth,angle=0]{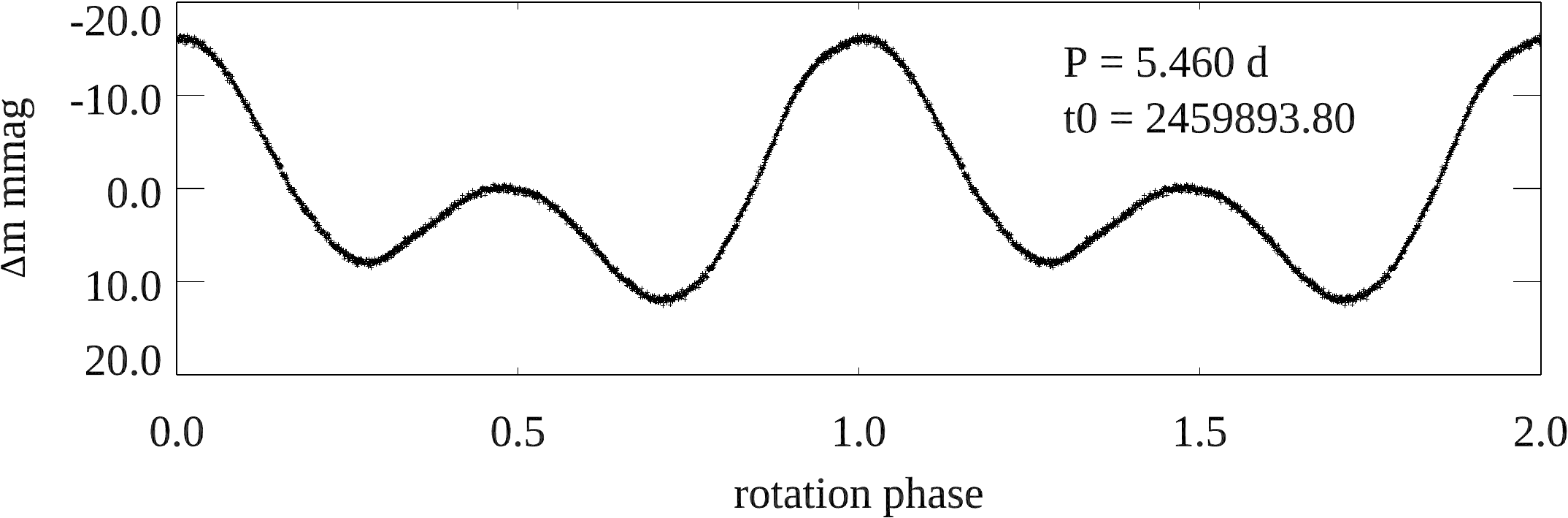}
\includegraphics[width=1.0\linewidth,angle=0]{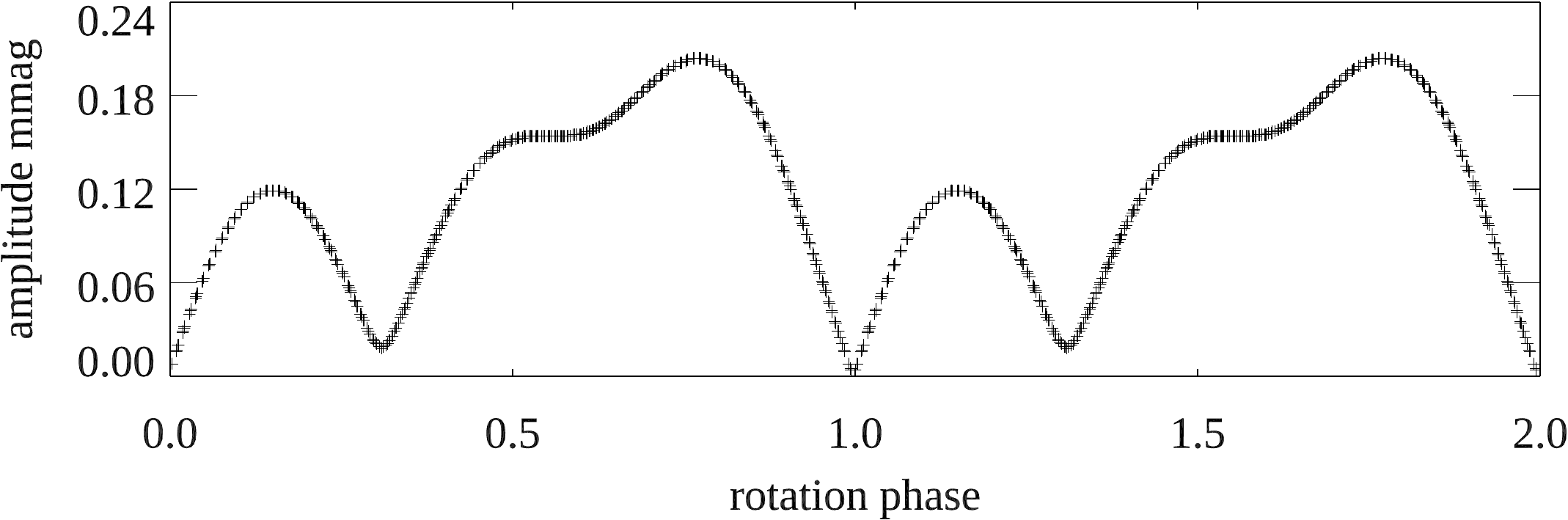}
\includegraphics[width=1.0\linewidth,angle=0]{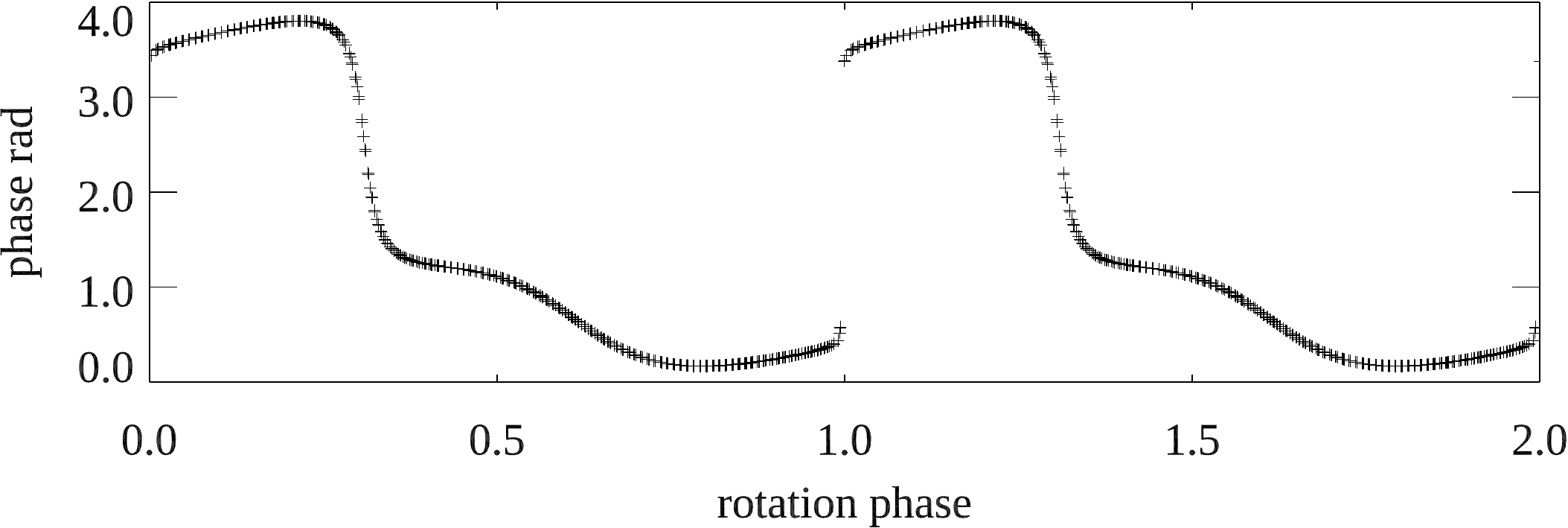}
\caption{Panel 1: The Sector 58 {\it TESS} light curve phased; the data have been binned by 10. Panels 2 and 3: The pulsation amplitude and phase as a function of rotation phase. These have been calculated from the frequency quadruplet only, with $\nu_2 = 183.34905$\,d$^{-1}$ as the mode frequency.  } 
\label{fig:phase1} 
\end{center}
\end{figure}

Several characteristics of the curves in Fig.\,\ref{fig:phase1} are notable. 1) The pulsation phase changes by $\uppi$ radians at the times when the pulsation amplitude goes to zero (or nearly so). This is characteristic of an oblique dipole pulsation when the pulsation node crosses the line of sight. 2) One pulsation pole dominates for about 30 per~cent of the rotation period and the other for about 70 per~cent. This is consistent with the longitudinal magnetic curve of  \citet{2005A&A...429L..55R}, which shows the longitudinal magnetic field to be negative for about 30 per~cent of the rotation and positive for the other 70 per~cent. 3) The double-humped pulsation amplitude maximum between rotation phases $0.3 -1.0$ shows that the dipole mode is heavily distorted on this magnetic hemisphere of the star. The corresponding pulsation phase is also significantly distorted. 4) The pulsation amplitude and phase variation on the other magnetic hemisphere does not appear to be distorted. 5) The time of brightest light in the rotation curve coincides with the time of one pulsation minimum. This suggests that the star is brightest around the pulsation equator and dimmest near the poles, as is characteristic of $\alpha^2$\,CVn stars.\footnote{The rotational photometric maximum and minimum in Ap stars is a function of the bandpass used for the observations. For some stars the maximum in a blue bandpass coincides with the minimum in a red bandpass. This is a consequence of the observations sampling different atmospheric depths and the line blocking from enhanced abundances changing the temperature gradient.} However, the pulsation mode is so distorted from a simple dipole, that this is a conjecture. 6) The mode frequency separation $\nu_6 - \nu_5 = 2.54$\,d$^{-1}$ ($29$\,$\upmu$Hz) is plausibly half the large separation, suggesting that these two modes are of alternatively even and odd degree, $\ell$, and that the large separation is about $\Delta \nu_0 \sim 60$\,$\upmu$Hz. 7) The other mode frequency separation $\nu_5 - \nu_2 = 13.350$\,d$^{-1}$ ($155$\,$\upmu$Hz) is then about $2.5\,\Delta \nu_0$, so that some intermediary modes are not excited to observable amplitude. This is seen in some other roAp stars, particularly, e.g., HD~217522 \citep{2015MNRAS.446.1347M}.

The rotational variations in the mean longitudinal magnetic field shown in Figure~2 of  \citet{2005A&A...429L..55R}  shows that pulsation maximum occurs at the time of positive magnetic extremum. The magnetic ephemeris of \citet{2005A&A...429L..55R} cannot be meaningfully extrapolated across the nearly 20-yr time gap between the magnetic measurements and the {\it TESS} photometry to compare the time of pulsation maximum in the {\it TESS} data with then extrema of the magnetic field. Interestingly, Figure~2 of \citet{2005A&A...429L..55R} shows that the mean longitudinal magnetic field of HD~12098 is negative for 30~per~cent of the rotation period and positive for 70~per~cent. Those are the same percentages of the rotation period for which we see the two pulsation amplitude maxima in Fig.\,\ref{fig:phase1}. We therefore conclude that the first pulsation maximum occurs when the negative magnetic hemisphere is observed, and the second,  more complex maximum  occurs when the positive magnetic hemisphere is visible. As with other roAp stars, it is likely that the pulsation amplitude maxima occur at the times of magnetic extrema, as appears to be the case in Figure~2 of \citet{2005A&A...429L..55R} for the Johnson $B$ observations of  \citet{2001A&A...380..142G} (see the next section). New contemporaneous magnetic and photometric observations are needed to confirm these reasonable conclusions.

There are no systematic studies of the pulsation amplitude and phase as a function of rotation phase, as shown in Fig.\,\ref{fig:phase1}, compared to the rotational light curves in roAp stars. With {\it TESS} data, such a study would be useful.

 \begin{table}
\centering
\caption{A linear least-squares fit of the frequencies derived from the Sector 58 data for HD~12098. The zero point for the phases is $t_0 = {\rm BJD}~2459893.80$, which matches the time of magnetic maximum  predicted from the ephemeris of  \citet{2005A&A...429L..55R}. The second frequency of the quadruplet, $\nu_2 = 183.3491$\,d$^{-1}$ ($2.1221$\,mHz) is taken to be the mode frequency from a distorted dipole.  The other two frequencies are assumed to be from independent pulsation modes, for which there is no sign of rotational sidelobes as expected for oblique pulsation; those, if present, could be lost in the noise. The frequencies $\nu_1$ to $\nu_4$ are over-specified to 5 decimals so that it can be seen that they are exactly equally split by $0.18315$\,d$^{-1}$ ($P_{\rm rot} = 5.460$\,d). }
\begin{tabular}{llcrr}
\hline
\hline
&\multicolumn{1}{c}{frequency} & \multicolumn{1}{c}{amplitude} &   
\multicolumn{1}{c}{phase} \\
&\multicolumn{1}{c}{d$^{-1}$} & \multicolumn{1}{c}{mmag} &   
\multicolumn{1}{c}{radians} \\
& & \multicolumn{1}{c}{$\pm 0.007$} & 
   \\
\hline
$\nu_1$ & $183.16590 \pm 0.00141 $ & $0.098 $ & $0.190 \pm 0.070 $  \\
$\nu_2$ & $183.34905 $ & $0.061 $ & $2.303 \pm 0.114 $  \\
$\nu_3$ & $183.53220 $ & $0.047 $ & $-2.056 \pm 0.148 $  \\
$\nu_4$ & $183.71535 $ & $0.040 $ & $-2.443 \pm 0.173 $  \\
$\nu_5$ & $196.6986 \pm 0.0040 $ & $0.035 $ & $1.470 \pm 0.200 $  \\
$\nu_6$ & $199.2402 \pm 0.0025 $ & $0.054 $ & $-1.832 \pm 0.128 $  \\
\hline\hline
\end{tabular}
\label{Tab:ls}
\end{table}

 \subsection{Comparison with ground-based data}
 
Observing through a Johnson $B$ filter, \citet{2001A&A...380..142G} obtained ground-based observations and found four pulsation frequencies in the range $187.7 - 198.7$\,d$^{-1}$ ($2.17 - 2.30$\,mHz). It is clear from comparison with Fig.\,\ref{fig:lc-ft2} that peaks in this frequency range are not present in the {\it TESS} Sector 58 data, although $\nu_6 = 199.2402$\,d$^{-1}$ ($2.3060$\,mHz) is close. Similarly, the frequency quadruplet seen in Fig.\,\ref{fig:lc-ft2} in the range $183.166 - 183.715$\,d$^{-1}$ ($2.120 - 2.126$\,mHz) is clearly not present in the \citet{2001A&A...380..142G} data (see Fig.\,\ref{fig:freq_model} in Section 3 below). As some roAp stars are known to have modes that show amplitude variation on time scales as short as days, and others have been observed to change modes on longer time scales, this may be the explanation for the different frequency ranges found in these two studies. 
 
Alternatively, it is known that the pulsation amplitudes and phases in roAp stars vary strongly both as a function of atmospheric depth and over the surface as seen in the spectral lines of elements trapped in abundance spots, usually associated with the magnetic poles (see, e.g., \citet{2006ESASP.624E..33K} and \citet{2009MNRAS.396..325F} for graphic examples). A theoretical study of magneto-acoustic modes as a function of atmospheric depth by \citet{2018MNRAS.480.1676Q} provides insight into how the acoustic and Alfv\'en components of the modes vary throughout the line-forming layers of the observable atmosphere. It thus seems possible that mode observed by \citet{2001A&A...380..142G} in Johnson $B$ may have lower amplitudes, or be undetectable, in the deeper region observed by {\it TESS} with its red bandpass. 

Yet unexplained complexity in the roAp pulsation modes and in the application of the oblique pulsator model to those stars gives rise to uncertainty in the inference of geometrical information about those modes. Two examples are the suggested discovery of modes with two different pulsation axes in the roAp star KIC~10195926 \citep{2011MNRAS.414.2550K}, and the finding that the pulsation geometry inferred from application of the oblique pulsator model is very different in ground-based Johnson $B$ data and {\it TESS} red data in the roAp star HD~6532 \citep{2020ASSP...57..313K}.

The measured photometric amplitudes of pulsation modes in roAp stars vary significantly depending on the bandpass used to make the observations \citep{1998MNRAS.299..371M} with amplitudes dropping from the blue to the red. This is simply a consequence of the spectral energy distribution and the photometric amplitude being primarily the result of temperature variations. \citet{2019MNRAS.487.3523C} compared photometric amplitudes measured through Johnson $B$ and {\it TESS} red bandpass for some roAp stars and found that measurements in Johnson $B$ typically show about 6 times the amplitude of measurements in {\it TESS} red. However, there is also an atmospheric depth effect, since photometric observations in different bandpasses sample different atmospheric depths, on average. Given the strong dependence of pulsation amplitude on atmospheric depth, it is not possible to conclude whether HD~12098 has changed modes between the time of the ground-based Johnson $B$ observations and the {\it TESS} observations, or whether the detected modes have very different amplitudes when observed through different bandpasses.  Simultaneous observations in multiple bandpasses, including Johnson $B$, and with {\it TESS} are needed to discriminate between these possibilities.

\section{Modelling}

 \begin{table}
\centering
\caption{Adopted parameters of HD\,12098}
\begin{tabular}{lr}
\hline
\hline
Parallax $= 6.833\pm0.022$\,mas, {$T_{\rm eff}=7600$~K} & $^{1)}$\\
\hline
$V= 7.97$\,mag, ~$E(B-V)= 0.~$,  $BC= 0.~$ & $^{2)}$\\ 
\hline
$M_V=2.14$\,mag,  $\log(L/{\rm L}_\odot) = 1.04$,  { $\log T_{\rm eff}= 3.881$} & $^{3)}$\\
\hline
$^{1)}$ { GAIA DR3:} \citet{2023DR3,2016GAIA}\\
$^{2)}$ \citet{Netopil_2008}\\
$^{3)}$ $\log$ means logarithm of base 10. 
\end{tabular}
\label{Tab:param}

\end{table}

In this section we model the pulsational amplitude and phase modulations against rotational phase of the main p mode pulsation of HD\,12098 (middle and bottom panels of Fig.\,\ref{fig:phase1}). We obtain these modulations by integrating the eigenfunction of the p mode on the visible hemisphere at each rotational phase for an assumed set of angles $\beta$ and $i$; the angle between rotational/magnetic axes and the inclination angle of the line-of-sight against the rotation axis, respectively \citep{2004SaioGautschy,2016Holdsworth_K2}. We assume the limb-darkening parameter $\mu = 0.6$ in this paper.  

The pulsational eigenfunctions are calculated by taking into account the effect of a dipole magnetic field \citep{2005MNRAS.360.1022S} with strength specified by the field strength $B_{\rm p}$ at the magnetic poles. Non-adiabatic pulsation variables (assumed to be axisymmetric to the magnetic axis neglecting rotation effects) are represented as a sum of the terms proportional to the Legendre function $P_{\ell_j}(\cos \theta)$ with $\ell_j=\ell_0 + 2j$ where $j= 1, 2, \ldots j_{\rm max}$ ($\ell_0 = 0$ for even modes and $\ell_0=1$ for odd modes). We set $j_{\rm max} = 12$ in this study.

Although even and odd modes are independent of each other, there is no pure dipole, or quadrupole mode since the pulsation energy is distributed among other values of $\ell_j$ because of the effects of the magnetic field. However, for convenience, we call a mode a distorted dipole mode if the $\ell_j=1$ component is dominant, or a distorted quadrupole mode if the $\ell_j=2$ component is dominant.

First, we choose a stellar model for the pulsation analysis. Table~\ref{Tab:param} lists observational parameters of HD~12098 adopted from the literature from which we also calculate the bolometric luminosity. Taking into account these parameters, we chose a $M=1.75$\,M$_\odot$ model having $\log (L/{\rm L}_\odot) = { 1.034}$ and  $\log T_{\rm {eff}}({\rm K})={ 3.875}$ from the evolutionary models computed with the initial composition $(X,Z)=(0.70,0.02)$ in the fully ionised layers. The evolutionary models, common to our previous works \citep[e.g.][]{2016Holdsworth_K2,2018MNRAS.476..601H,2021Shi}, were based on assumptions similar to the polar model of \citet{2001Balmforth}, in which the helium mass fraction is depleted to 0.01 in the layers above the second He ionisation zone to the surface, and convection in the envelope is neglected assuming a strong magnetic field to stabilise the outer layers. 

 \begin{figure}
\begin{center}
\includegraphics[width=1.0\linewidth,angle=0] {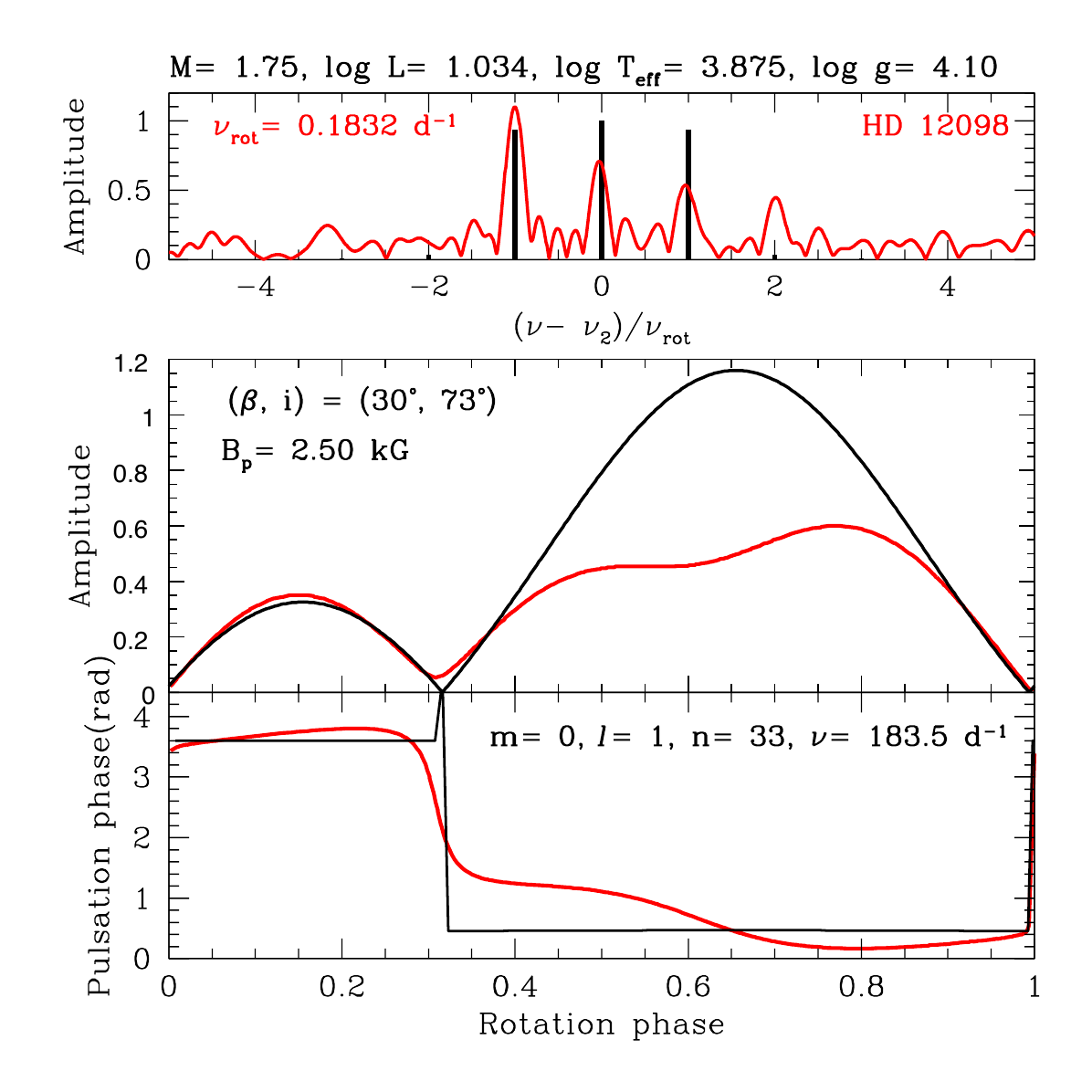}
\caption{Comparison of observed Fourier amplitude (top), rotational modulations of the pulsation amplitude (middle) and the phase modulation (bottom) of HD~12098 with a $1.75$\,M$_\odot$ dipole pulsator model at a 2.5\,kG magnetic field.} 
\label{fig:pul_model} 
\end{center}
\end{figure}

Fig.~\ref{fig:pul_model} compares a dipole pulsation model ($\ell_0 =1$) with the observed pulsational amplitude and phase modulations of HD~12098. For this model we have chosen angles $\beta=30^\circ$ and $i=73^\circ$ to match the rotation phase at which sudden changes of the pulsation phases are seen. We have normalised the model amplitude to approximately fit the (secondary) maximum at a rotation phase of 0.15. At this phase the rotation axis, the line-of-sight and the magnetic axis are on a meridional plane with the visible magnetic pole at the angle $(180^\circ-\beta-i)$ from the line of sight. During the range of rotation phase between 0.0 to 0.3 (when the negative magnetic pole is seen in the lower part of the visible hemisphere) the amplitude and phase variations reasonably agree with those observed. 

During the rotation phases between 0.3 and 1.0, the positive magnetic pole is visible. The angle between the visible pole and the line of sight attains minimum $(i-\beta)$ at the rotation phase 0.65, when the predicted pulsation amplitude is maximum.  However, the predicted amplitude during this range of rotation phase deviates considerably from the observed one. The observed amplitude variation is asymmetric with a bump and tends to be lower than the model prediction. In this range of rotation phase, the other positive magnetic pole is visible with angles between $(i-\beta)$ and ${\pi\over2}$\,rad from the line of sight.  The cause of the deviation, and in particular the cause of the broad bump in amplitude, is not clear. The pulsation phase of our model is constant during this phase interval, while the observed one gradually decreases, which is probably caused by the effect of rotation (neglected in this model) as discussed in \citet{2011A&A...536A..73B}.

The deviation of the predicted amplitude variation from the observed one results in a lack of the fourth frequency component in the Fourier spectrum shown in the top panel of Fig.~\ref{fig:pul_model}. Our model always predicts a triplet-dominated frequency spectrum for an odd-mode oscillation even if it is significantly affected by contributions from components with $\ell_j \ge 3$, because such contributions are mostly cancelled through the surface integration as discussed in \citet{2004SaioGautschy}. For this reason, the Fourier amplitude at { $(\nu-\nu_2)/\nu_{\rm rot} = 2$} detected in the pulsation of HD~12098 cannot be explained by an axisymmetric odd mode. Some non-axisymmetric variation is needed for the Fourier component. In this respect, the bump in the amplitude modulation of HD~12098 might be caused by a non-axisymmetric (with respect to the magnetic axis) phenomenon and might be related with the Fourier component at { $(\nu-\nu_2)/\nu_{\rm rot} = 2$}.

\begin{figure}
\begin{center}
\includegraphics[width=1.0\linewidth,angle=0] {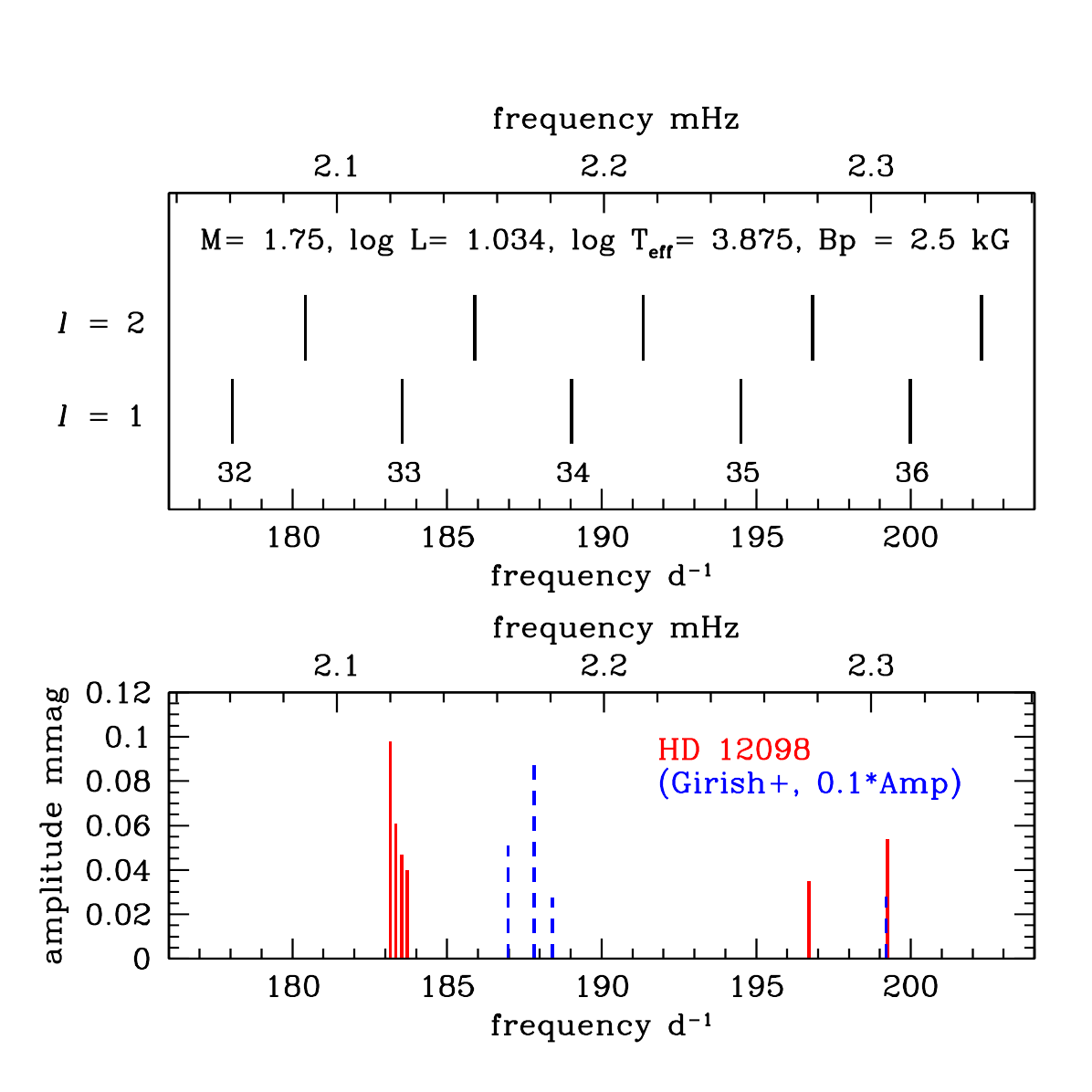}
\caption{Comparison of observed frequencies with dipole and quadrupole modes of our model at $B_{\rm p}=2.5$~kG. The integer along each frequency of the dipole $(\ell = 1)$ modes indicates the radial order. { The blue dashed lines show the frequencies found by   \citet{2001A&A...380..142G} from ground-based Johnson $B$ data.} } 
\label{fig:freq_model} 
\end{center}
\end{figure}

Our model has a p-mode large separation of  $5.47~$d$^{-1}($63.3$\,\mu$Hz) similar to the observational value $\sim$$60$\,$\mu$Hz. Fig.\,\ref{fig:freq_model} compares model frequencies of distorted dipole and quadrupole p modes at $B_{\rm p}=2.5$\,kG (upper panel) in the frequency range of HD~12098 (lower panel). This figure indicates that the observed highest frequency $\nu_6$ corresponds to the dipole mode of order 36, while $\nu_5$ is the quadrupole mode of order 35. In addition, it is interesting to note that the frequency $189.2$~d$^{-1}$ (2.19~mHz) obtained by \citet{2000IBVS.4853....1M} is close to a dipole mode of order 34, one order higher than the order 33 for $\nu_2$, indicating that the most strongly excited modes seem to shift with time, { or that different modes are detected at different atmospheric depths (see section 2.4 above). We do not know why the frequencies of highest amplitude found by  \citet{2001A&A...380..142G}  do not match any of our model mode frequencies}. We note, however, that these frequencies are all above the acoustic cut-off frequency, $\approx 135~$d$^{-1}$ ($1.56$\,mHz), { as in a number of other roAp stars (see Figure\,12 of \citealt{2018MNRAS.476..601H})}. The excitation mechanism for these high-frequency pulsations in roAp stars is not known.

Finally, we find that quadrupole modes (even modes ($\ell_0=0$) with the maximum amplitude at $\ell_j = 2$) affected with similar magnetic fields are not good for HD~12098. This is because the pulsational phase variations are significantly suppressed just as in our previous cases presented in, e.g., \citet{2018MNRAS.476..601H} and \citet{2016Holdsworth_K2}.     

\section{conclusions} 

HD~12098 is an roAp star with a dipolar magnetic field and a rotation period of $P_{\rm rot} = 5.460$\,d. Using data from Sector 58 of the {\it TESS} mission, we have shown that there are rotational light variations consistent with spots, or patches, of enhanced abundances near to each of the magnetic poles. In these senses, HD~12098 is a typical $\alpha^2$\,CVn star. 

HD~12098 is a known roAp star \citep{2001A&A...380..142G}  for which the {\it TESS} data give far more insight to the pulsations than previous ground-based data. The star pulsates with several modes in the range $183 - 200$\,d$^{-1}$ ($2.12 - 2.31$\,mHz). While this encompasses the same frequency range found by \citeauthor{2001A&A...380..142G}, the star has either changed modes, or the observations through the Johnson $B$ bandpass and the {\it TESS} red bandpass sample different depths with different mode visibility. To discriminate between these possibilities requires simultaneous Johnson $B$ and {\it TESS} observations. 

The interesting new discovery for HD~12098 is that its principal pulsation mode is a dipole mode that is far more distorted than has been observed in other roAp stars. Using models that have reasonably explained the mode distortion in other roAp stars, we are unable to account for the double-humped pulsation amplitude modulation between rotation phases $0.3 - 1.0$. This characteristic has not been observed in any other roAp star. On the dipole pulsation mode hemisphere seen with best aspect during those $0.3 - 1.0$ rotation phases, the mode is strongly distorted from a simple dipole. 

Interestingly, recently oblique pulsators have been discovered in close binary stars where dipole pulsations are strongly trapped in one hemisphere, or another -- the tidally tilted pulsators and so-called single-sided pulsators (\citealt{2020NatAs...4..684H}, \citealt{2020MNRAS.494.5118K}, \citealt{2020MNRAS.498.5730F},  \citealt{2021MNRAS.503..254R}, \citealt{2023arXiv231116248Z}). Whether the clear asymmetry of the pulsation in the dipole hemispheres of HD~12098 has any relation to this is unknown. 

HD~12098 also shows a low-overtone p~mode and possibly some g~modes. These, too, are unusual in roAp stars and theoretically unexpected \citep{2005MNRAS.360.1022S} for this star's magnetic field strength of over a kG \citep{2005A&A...429L..55R}. This star calls for simultaneous multi-colour photometric observations, and it presents new, challenging behaviour to theory.

\section*{acknowledgements}
DLH and DWK acknowledge support from the Funda\c c\~ao para a Ci\^encia e a Tecnologia (FCT) through national funds (2022.03993.PTDC). This paper includes data collected by the {\it TESS} mission. Funding for {\it TESS} is provided by NASA's Science Mission Directorate. Resources used in this work were provided by the NASA High End Computing (HEC) Program through the NASA Advanced Supercomputing (NAS) Division at Ames Research Center for the production of the SPOC data products. 

\section*{data availability}

The {\it TESS} data used in this study are available on MAST.

\bibliographystyle{mnras}
\bibliography{hd12098}{}
\end{document}